\documentclass[prl,superscriptaddress,reprint]{revtex4-1}
\usepackage{graphicx}
\usepackage{dcolumn}
\usepackage{bm}
\usepackage[latin1]{inputenc}

\begin{document}

\title{Spin Wave Diffraction and Perfect Imaging of a Grating}

\author{S. Mansfeld}
\author{J. Topp}
\author{K. Martens}
\author{J. N. Toedt}
\author{W. Hansen}
\author{D. Heitmann}
\author{S. Mendach}\email{smendach@physnet.uni-hamburg.de}

\affiliation{%
Institut f\"ur Angewandte Physik und
Mikrostrukturforschungszentrum, Universit\"at Hamburg,
Jungiusstrasse 11, D-20355 Hamburg, Germany}%
\date{\today}

\begin{abstract}
We study the diffraction of Damon-Eshbach-type spin waves incident
on a one-dimensional grating realized by micro slits in a thin
permalloy film. By means of time-resolved scanning Kerr microscopy
we observe unique diffraction patterns behind the grating which
exhibit replications of the spin-wave field at the slits. We show
that these spin-wave images, with details finer than the wavelength
of the incident Damon-Eshbach spin wavelength, arise from the
strongly anisotropic spin wave dispersion.
\end{abstract}
\maketitle
The possibility to use spin waves to excite, store and retrieve
electric signals and to perform logical operations relies on the
ability to manipulate spin wave propagation. In patterned, thin
ferromagnetic films spin waves with gigahertz frequencies and
wavelengths from several hundreds of nanometers up to millimeters
can be
confined~\cite{Jorzick1999,Buess2004,Bayer2005,Perzlmaier2005,Neudecker2006a,Podbielski2006,Mendach2008,Balhorn2010},
focused and
guided~\cite{Hillebrands1997,Bauer1998,Demidov2007,Topp2008,Demidov2008b,
Demidov2009c}. An intriguing possibility to guide spin waves in
unpatterned films, lies in their anisotropic dispersion. The
dispersion of dipole-dominated spin waves in thin ferromagnetic
films strongly depends on their propagation direction~$\vec{k}$ with
respect to an in-plane magnetic field~$\vec{H}_0$ giving rise to
Damon-Eshbach modes with $\vec{k}_{DE}\perp\vec{H}_0$,
Backward-Volume modes with $\vec{k}_{BV}\parallel\vec{H}_0$, or a
combination of both. This anisotropy in the dispersion of spin
waves~\cite{Buettner2000,Demidov2005,Perzlmaier2008} has recently
been studied theoretically~\cite{Veerakumar2006} and
experimentally~\cite{Demidov2009d,Schneider2010,Kostylev2011} as a
novel way to manipulate spin wave propagation.

In this letter we demonstrate that the anisotropic spin-wave
dispersion enables perfect imaging with spin waves. We present and
discuss time resolved scanning Kerr microscopy
(TR-SKM)~\cite{Freeman1996, Tamaru2004} data on the diffraction of
planar Damon-Eshbach spin waves on a one-dimensional grating
realized by micrometer sized slits in a permalloy film. Behind this
grating, we observe a unique diffraction pattern which arises from
the interference of Damon-Eshbach and Backward-Volume type spin
waves and their inherent strongly anisotropic dispersion. We show
that this diffraction pattern produces images of the spin wave field
at the slits. The resolution of these images is not limited by the
wavelength of the incident spin wave, as it is the case in isotropic
media~\cite{Abbe1873}, but deviations from a perfect image occur
solely due to spin wave damping and due to the finite curvature of
the dispersion's iso-frequency line in $k$ space. In that sense, our
spin wave images represent a spin-wave analogue to sub-wavelength
resolution concepts with anisotropic media in optical
metamaterials~\cite{Liu2007b, Schwaiger2009}. In addition, we
demonstrate that the position of the spin wave images behind the
slit array can be tuned by manipulating the spin-wave dispersion via
the excitation frequency and the external magnetic field. This
enables us to create tailor-made spin-wave fields in an unpatterned
ferromagnet film with sub-micron resolution.

\begin{figure}
\begin{center}
\includegraphics[width= 75 mm]{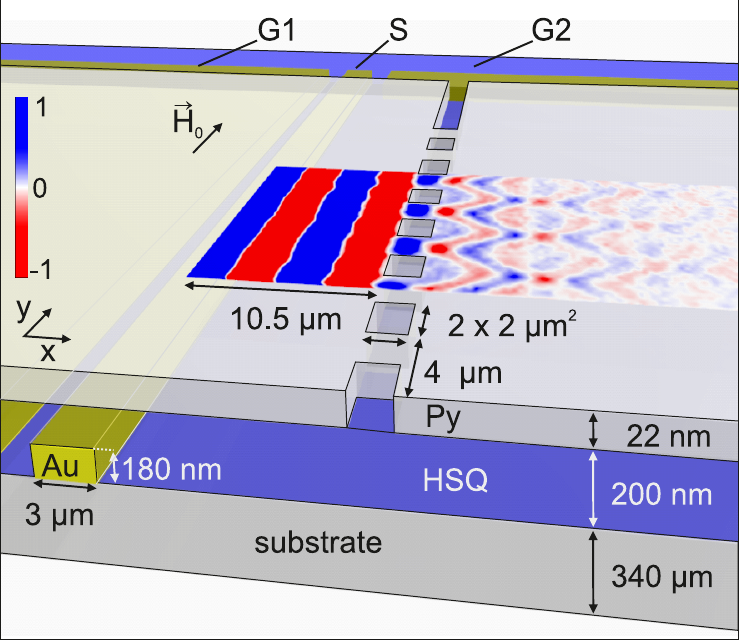}
\caption{FIG. 1 (color online): Sketch of the experimental arrangement. S:
Signal line, G1, G2: Ground lines of the coplanar wave guide. An
exemplary TR-SKM phase image of the spin wave field taken at
$\mu_{0} H_{0} = 10$~mT and $f=4180$~MHz is overlaid as a color
density plot on a part of the Permalloy film.}\label{Figure1}
\end{center}
\end{figure}

The experimental arrangement is sketched in Fig.~1. The investigated
sample is a 22~nm thick Permalloy film prepared on top of a coplanar
wave guide (CPW) structure. The CPW was prepared on a $340$~$\mu$m
thick GaAs(100) substrate and consists of a $3$~nm thick adhesive
layer of chromium and a $180$~nm thick layer of gold. It features a
3~$\mu$m wide signal line~(S) separated by 2~$\mu$m gaps from the
ground lines (G1,G2). G2 exhibits a $150$~$\mu$m wide gap at the
position of the Permalloy film; this guarantees the injection of
well defined planar spin waves from the signal line~(S). An
approximately $200$~nm thick layer of hydrogen silsesquioxane was
applied on top of the wave guide acting as an insulator between the
wave guide and the Permalloy film. The film has a rectangular shape
with a length of \mbox{$l = 150$ $\mu$m} perpendicular and $w = 120$
$\mu$m parallel to the CPW. At a distance of $10.5$~$\mu$m from the
center of the signal line~(S), the Permalloy film features an
one-dimensional grating realized by $13$ square-shaped holes with a
side length of $2$ $\mu$m and a periodicity of $p=4$ $\mu$m. The
permalloy film was prepared using vacuum thermal evaporation in
combination with electron beam lithography and lift off technique.

TR-SKM is used to examine the spin wave propagation in the Permalloy
film. A pulsed Titan-Sapphire laser with a wavelength of $800$~nm
and a repetition rate of 76~MHz is focused onto the sample to probe
the magnetization component perpendicular to the Permalloy film via
the magneto optical Kerr effect. The laser focus has a width of
approximately $550$~nm and is scanned over the sample in $250$~nm
steps. Spin waves are excited in the film by passing a continuous
microwave through the CPW. The customized microwave synthesizer (ITS
9200) used to excite the spin waves is phase locked to the
repetition rate of the laser. By tuning the phase offset between
laser and microwave synthesizer, the spin-wave field in the sample
can be measured phase dependently. This enables us to image
propagating spin wave fields with a temporal resolution of a few
Picoseconds. Initially, we have optimized the set up shown in Fig.~1
with experiments on homogenous unpatterned films and found that,
except for edge effects which extent less than 15~$\mu$m from the
edges of the film, the CPW emits planar spin waves with wave vectors
tuneable up to $k = 4.5$ $\mu$$m^{-1}$. The color encoded plot
overlaid onto the sample sketch in Fig.~1 represents an exemplary
measurement. As indicated by the arrow, the external magnetic field
is oriented parallel to the CPW. All magnetic fields quoted in the
following have been applied after applying a saturation field of
80~mT in y direction and ramping down to 0~mT. A Damon-Eshbach type
spin wave is excited at the signal line~(S) and propagates towards
the grating. At the grating the spin wave is partly reflected and
partly transmitted. In the following we investigate the transmitted
part and its unique interference patterns. We show that these
patterns produce images of the grating at distinct positions.

\begin{figure}
\begin{center}
\includegraphics[width= 85 mm]{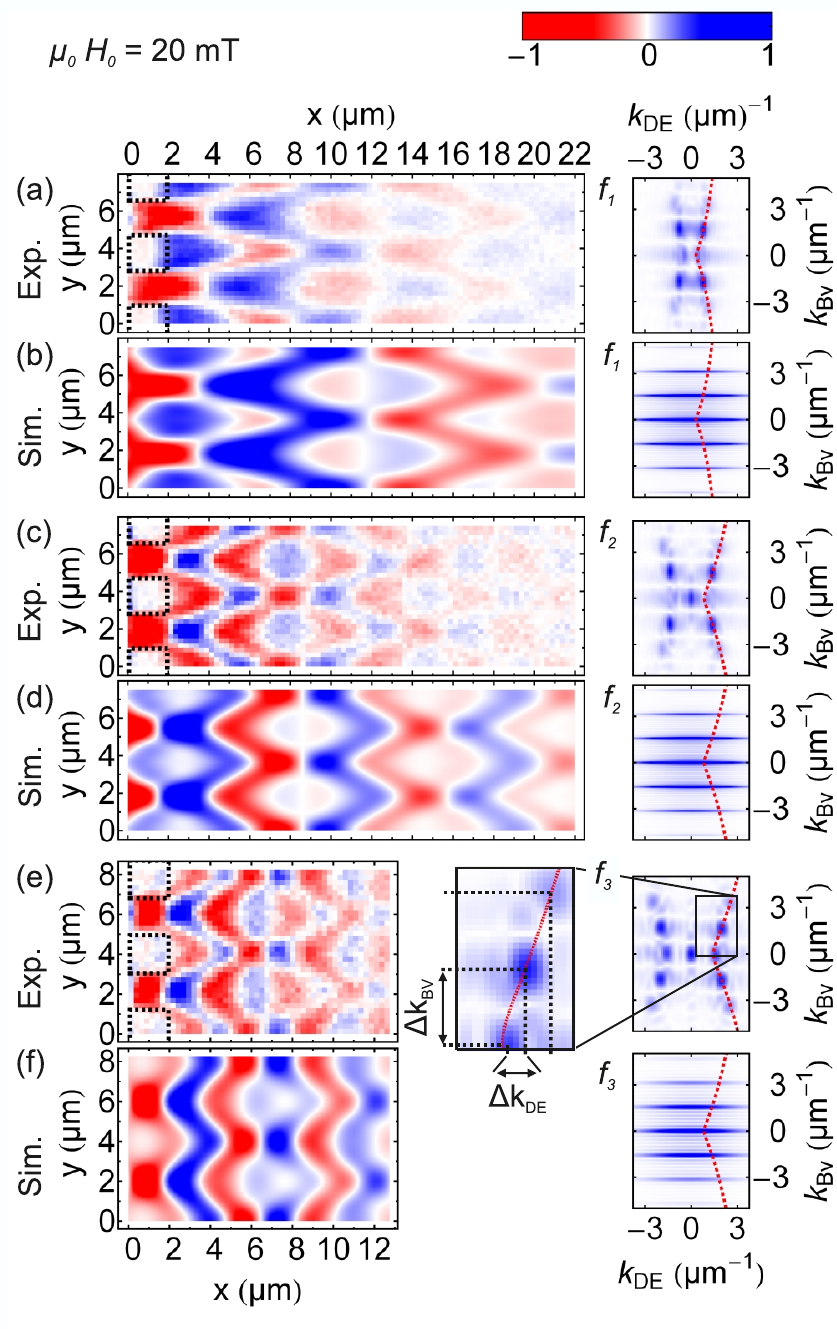}
\caption{FIG. 2 (color online): (a), (c) and (e) TR-SKM phase image of the
spin-wave field (left side) behind the grating and corresponding
Fourier spectra (right side) measured at $H_{0} = 20$~mT with $f_{1}
= 4180$~MHz, $f_{2} = 4636$~MHz, and $f_{3} = 5092$~MHz,
respectively. The holes in the Py-film defining the grating are
marked by dashed black lines in the TR-SKM images on the left side.
(b), (d) and (f) Calculated spin-wave fields for $f_{1} = 4180$~MHz,
$f_{2} = 4636$~MHz and $f_{3} = 5092$~MHz (left side) and Fourier
spectra of an analytical approximation of the slit array. Here
$k_{DE}$ and $k_{BV}$ are, respectively, spin-wave vector components
in x direction and in y direction. The red dashed lines
plotted in all Fourier spectra mark the theoretical DIFL.} \label{Figure2}
\end{center}
\end{figure}

Let us first concentrate on three TR-SKM images depicted on the left
hand sides of Figs. 2(a),(c),(e) and measured at $\mu_0 H_0 = 20$~mT
with an excitation frequency of $f_{1} = 4180$~MHz, $f_{2} =
4636$~MHz and $f_{3} = 5092$~MHz, respectively. These images show
the propagating spin wave field at a defined phase of the
excitation, i.e., they represent snap shots of the spin wave field
for a certain point in time. In the following they are referred to
as phase plots. The holes in the Py-film defining the grating are
marked by dashed black lines. For all frequencies we observe
distinct interference patterns which are periodic in propagation
direction ($x$ direction) and along the grating ($y$ direction).
With increasing frequency we find that the period in propagation
direction gets smaller while the period along the grating is fixed.
This is also reflected in the corresponding spatial Fourier spectra
of the area behind the grating, which are shown on the right hand
side of Figs.~2(a),(c),(e). We observe well defined intensity maxima
in the Fourier spectra which correspond to the spin-wave modes
contributing to the respective interference pattern. Additionally,
we find peaks for $k_{DE} = 0$ and $k_{BV} = 0$, which are not
related to the propagating spin-wave field. The Backward-Volume
components $\left|k_{BV}\right|$ of the observed modes correspond to
the grating period in $y$ direction ($p=4$ $\mu$m) and are the same
for all excitation frequencies ($\left|k_{BV0}\right| = 0$,
$\left|k_{BV1}\right| = 1.65~ \mu m^{-1}$, $\left|k_{BV2}\right| =
3.30~ \mu m^{-1}$). In contrast to this, the related Damon-Eshbach
components $\left|k_{DE}\right|$ which correspond to the period in
propagation direction increase with increasing frequency. They are
imposed by the anisotropic dispersion law of spin waves in the Py
film, since only wave vectors fullfilling this dispersion law can
propagate away from the grating. Therefore, the experimental Fourier
spectra shown in Figs.~2(a),(c),(e) give the convolution of the
grating's Fourier spectrum with the dispersion's iso-frequency line
in $k$ space~(DIFL) at different frequencies (The DIFL is also
referred to as `slowness curve' in recent
literature~\cite{Veerakumar2006,Kostylev2011}). Indeed, we find that
the observed maxima fit very well with calculated DIFLs (red dashed
lines) obtained with an analytical model by Guslienko et
al.~\cite{Guslienko2003} for a 22~nm thin ferromagnetic film with a
saturation magnetization of $\mu_{0} M_{s} =0.94$~mT.

A first-principal calculation of the diffraction pattern would be an
extremely challenging task. It would require a detailed knowledge of
the static spin boundary condition at the edges of the holes and the
arising static and dynamic magnetization pattern at the grating in
an external magnetic field. However, we found that with a simple
analytical ansatz we can well reproduce the observed interference
pattern and get a deep insight into the microscopic spin wave
propagation: We use an array of 13 intensity distributions
$s(\vec{r}) = S_{0}\cdot$Exp$(-2 \vec{r}^{~2}~\mu$m$^{-2})$ spaced
$4$~$\mu$m apart as an approximation of the spin-wave distribution
at the grating. Phase and amplitude of all modes excited by the
diffraction are calculated by taking the Fourier spectrum of this
intensity distribution. To obtain those modes of the spectrum which
can propagate in the Permalloy film, we convolute the Fourier
spectrum of the grating with the calculated DIFLs. This is
illustrated in the right hand sides of Figs.~2(b),(d),(f). By
superimposing all, in this way identified propagating modes, with
the correct amplitude and phase in real space, we calculate the
spin-wave interference patterns shown on the left hand sides of
Figs.~2(b),(d) and (f). We find a good agreement with the
corresponding experimental data shown in Figs.~2(a),(c),(e). Small
deviations in the details of the spin-wave patterns are due to the
simple approximation we use for the spin-wave distribution at the
slit array. Note that it is not sufficient to assume point sources
to obtain a good agreement between calculation and experiment.

\begin{figure}
\begin{center}
\includegraphics[width= 85 mm]{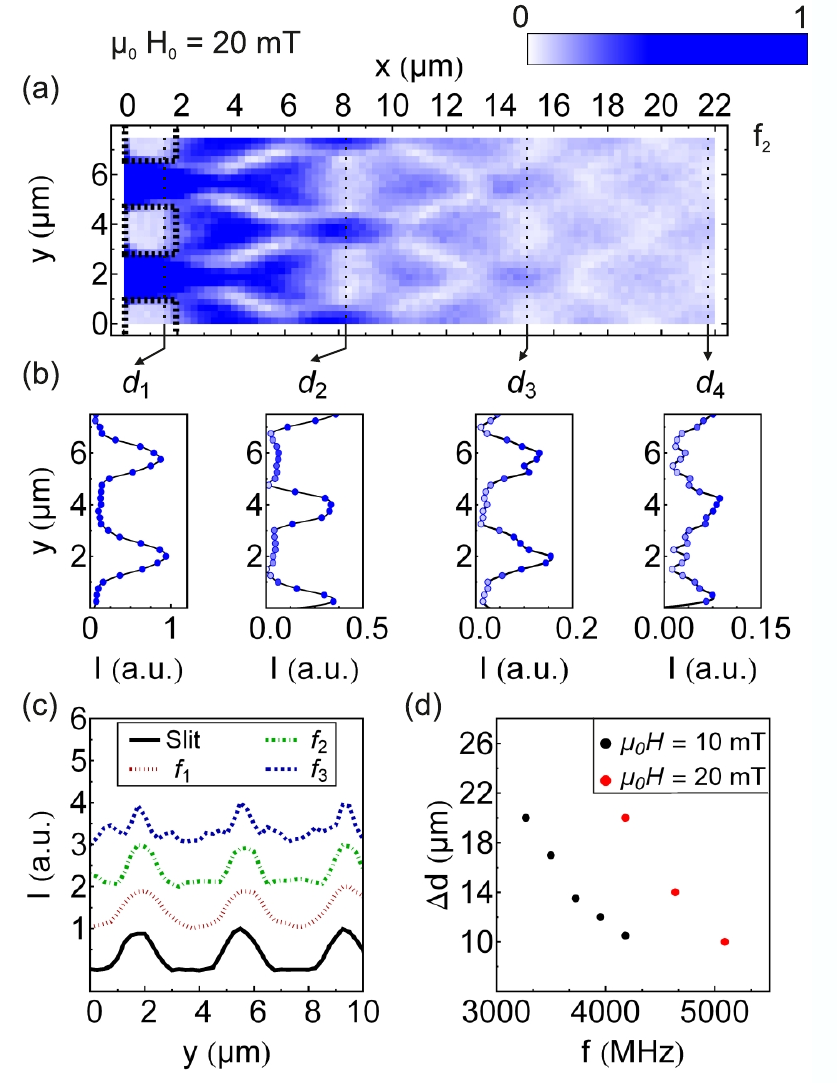}
\caption{FIG. 3 (color online): (a) TR-SKM amplitude image of the spin-wave
field at $H_{0} = 20$~mT and $f_{2} = 4636$~MHz corresponding to the
left side of Fig. 2(c). The holes in the Py film defining the
grating are marked by thick dashed black lines. (b) Cross sections
along the thin dashed black lines at $d_1$,$d_2$,$d_3$ and $d_4$.
(c) Cross sections along the first image line measured at  $f_{1} =
4180$~MHz, $f_{2} = 4636$~MHz and $f_{3} = 5092$~MHz. (d) Distance
between grating and image line $\Delta d$ at $\mu_{0} H_{0} = 10$~mT
(black dots) and $\mu_{0}
H_{0} = 20$~mT (red dots).}\label{Figure3}
\end{center}
\end{figure}

Figure~3(a) shows an amplitude plot which was measured at $f_{2} =
4636$~MHz and $H_{ext} = 20$~mT and corresponds to the phase plot
shown in Fig.~2(c). To obtain the amplitude plot, we scanned through
a complete phase period and measured the maximum deflection of the
magnetization for all spatial points in the measured field. We find
distinct areas with high amplitudes (blue color) and areas with
amplitudes near zero (white color). The sharp features in the
pattern, e.g. the white lines with a small width of approximately
700~nm compared to the incident Damon-Eshbach wavelength of
$6.3~\mu$m, indicate the excitation of high $k_x$ and $k_y$
components at the grating.

Most interestingly, the amplitude distribution at the
grating~($d_1=1.50~\mu$m) is replicated at equidistant $x$~values
$d_2=8.25~\mu$m, $d_3=15.00~\mu$m, and $d_4=21.75~\mu$m where every
other replication is shifted in $y$ direction by half of the grating
period~$p$. Figure~3(b) illustrates the replications at the
respective $x$~values. They represent spin-wave images of the
grating cross-section as will be explained in the following.

In general, to obtain a perfect image of the grating cross-section
(i) all Fourier components present at the grating cross-section have
to reach the respective image lines and (ii) they have to be
superimposed with the right phase and with the right weighting of
amplitudes. As visible from the Fourier spectra shown on the right
hand side of Figs.~2(c) and (d), condition~(i) is fulfilled because
the DIFL (dashed red line) crosses all grating diffraction orders
which exhibit a considerable intensity, i.e. all Fourier components
necessary to obtain an image of the grating cross-section can
propagate in our film. Condition~(ii) is fulfilled due to the almost
linear shape of the DIFL in this wave vector regime and frequency
regime. Due to this linearity, the spacing between Fourier
components in cross-sectional direction ($y$~direction), which is
given by the reciprocal lattice vector $\Delta k_{BV}=2\pi/p$ of the
grating, leads to an equidistant spacing of the Fourier components
in propagation direction with $\Delta k_{DE}=\gamma\Delta k_{BV}$
and $\gamma$ being the slope of the DIFL (cf. zoom in of the Fourier
spectrum in Fig.~2(e)). As a result, beating patterns form in real
space exhibiting image lines at periodic distances~$\Delta
d=2\pi/\Delta k_{DE}=p/\gamma$, at which the phases of all Fourier
components correspond to the grating cross-section. Furthermore, at
half way between these image lines additional image lines exist,
where the phases of all Fourier-components are shifted by
180$^{\circ}$ compared to the grating cross-section, leading to an
image shift in $y$ direction by $p$/2. Since the cross-section of
the spin-wave field at all $x$ positions inside the slits is imaged
as explained above, we in fact observe two-dimensional images of the
slits in Fig.~3(a).

Since the slope~$\gamma$ of the DIFL changes with frequency~$f$ (cf.
Figs.~2(a),(c),(e)) and magnetic field $H_0$, we can tailor the
period of the image lines~$\Delta d=p/\gamma$, i.e. their position
in the film, via $f$ or $H_0$. As shown in Fig.~3(d), we find that
$\Delta d$ increases with increasing magnetic field and decreasing
frequency. With the parameters of our experiment we obtain values
between 10~$\mu$m and 20~$\mu$m.

Figure~3(c) compares how the spin-wave cross-section at $d_1$ is
imaged in the respective image lines for $f_{1} = 4180$~MHz, $f_{2}
= 4636$~MHz and $f_{3} = 5092$~MHz at $\mu_0 H_0 = 20$~mT. While the
image quality at $f_1$ and $f_2$ is comparable, the shorter spin
wave damping length at $f_3$ leads to a reduction in image quality.
In measurements on a 20~nm thick Py film without grating we found
that the 1/e amplitude damping length decreases from approximately
17~$\mu$m to approximately 5~$\mu$m if we increase the excitation
frequency from 3800~MHz to 5000~MHz. Besides the spin wave damping,
which degrades the quality of the images also with increasing
distance from the grating (cf.~Fig.~3(b)), deviation from a perfect
image occur due to the finite curvature of the DIFL which smears out
the $x$ position of the image lines. Note that, as visible at
$d_2=8.25~\mu$m in Fig.~3(b), this smearing out can also lead to a
reduction of the imaged slit width compared to the slit width in the
object at $d_1=1.50~\mu$m.

Aside from the perfect spin-wave images presented here another
consequence of the linear DIFL is the formation of caustic spin wave
beams with direction given by the normal of the DIFL as discussed in
recent literature~\cite{Demidov2009d,Schneider2010,Kostylev2011}.
While these caustic beams do not appear in the diffraction pattern
of our grating structure, we indeed observe caustic beams in a
double-slit structure (not shown) which was prepared in accordance
to the grating structure. The appearance of caustic beams in the
double-slit structure is most probably due to its smoother Fourier
spectrum compared to the grating. Interestingly, we found one single
slit's spin wave image at the crossing of the caustic beams. This
reflects the fact that the crossing points of the caustic beams
coincide with the image lines discussed above.

In conclusion, we investigated the diffraction of planar
Damon-Eshbach spin waves incident on a grating and on a double slit
by means of TR-SKM measurements. We found that due to the linearly
shaped DIFL spin-wave images of the slits form behind the
structures. The quality of these images is only limited by spin wave
damping and the deviation of the DIFL from a linear shape.
Furthermore, we demonstrated that the image position in the film
behind the periodic structures can be tuned by changing the slope of
the DIFL via the excitation frequency or external magnetic field.
These findings enable new concepts for controlled spin-wave
confinement and manipulation in unpatterned ferromagnetic films via
adjacent slit arrays, e.g. to realize novel spin-wave filters or
logic devices.

We acknowledge major contributions to the setting up of our TR-SKM
by Jan Podbielski and Andreas Krohn as well as fruitful discussions
with Andreas Rottler and Stephan Schwaiger and financial support by
the DFG via SFB 668, SFB 508, GrK 1286, and by the City of Hamburg
via the Cluster of Excellence Nano-Spintronics.
\newpage


\begin{thebibliography}{27}%
\makeatletter
\providecommand \@ifxundefined [1]{%
 \@ifx{#1\undefined}
}%
\providecommand \@ifnum [1]{%
 \ifnum #1\expandafter \@firstoftwo
 \else \expandafter \@secondoftwo
 \fi
}%
\providecommand \@ifx [1]{%
 \ifx #1\expandafter \@firstoftwo
 \else \expandafter \@secondoftwo
 \fi
}%
\providecommand \natexlab [1]{#1}%
\providecommand \enquote  [1]{``#1''}%
\providecommand \bibnamefont  [1]{#1}%
\providecommand \bibfnamefont [1]{#1}%
\providecommand \citenamefont [1]{#1}%
\providecommand \href@noop [0]{\@secondoftwo}%
\providecommand \href [0]{\begingroup \@sanitize@url \@href}%
\providecommand \@href[1]{\@@startlink{#1}\@@href}%
\providecommand \@@href[1]{\endgroup#1\@@endlink}%
\providecommand \@sanitize@url [0]{\catcode `\\12\catcode
`\$12\catcode
  `\&12\catcode `\#12\catcode `\^12\catcode `\_12\catcode `\%12\relax}%
\providecommand \@@startlink[1]{}%
\providecommand \@@endlink[0]{}%
\providecommand \url  [0]{\begingroup\@sanitize@url \@url }%
\providecommand \@url [1]{\endgroup\@href {#1}{\urlprefix }}%
\providecommand \urlprefix  [0]{URL }%
\providecommand \Eprint [0]{\href }%
\providecommand \doibase [0]{http://dx.doi.org/}%
\providecommand \selectlanguage [0]{\@gobble}%
\providecommand \bibinfo  [0]{\@secondoftwo}%
\providecommand \bibfield  [0]{\@secondoftwo}%
\providecommand \translation [1]{[#1]}%
\providecommand \BibitemOpen [0]{}%
\providecommand \bibitemStop [0]{}%
\providecommand \bibitemNoStop [0]{.\EOS\space}%
\providecommand \EOS [0]{\spacefactor3000\relax}%
\providecommand \BibitemShut  [1]{\csname bibitem#1\endcsname}%
\let\auto@bib@innerbib\@empty
\bibitem [{\citenamefont {Jorzick}\ \emph {et~al.}(1999)\citenamefont
  {Jorzick}, \citenamefont {Demokritov}, \citenamefont {Mathieu}, \citenamefont
  {Hillebrands}, \citenamefont {Bartenlian}, \citenamefont {Chappert},
  \citenamefont {Rousseaux},\ and\ \citenamefont {Slavin}}]{Jorzick1999}%
  \BibitemOpen
  \bibfield  {author} {\bibinfo {author} {\bibfnamefont {J.}~\bibnamefont
  {Jorzick}}, \bibinfo {author} {\bibfnamefont {S.}~\bibnamefont {Demokritov}},
  \bibinfo {author} {\bibfnamefont {C.}~\bibnamefont {Mathieu}}, \bibinfo
  {author} {\bibfnamefont {B.}~\bibnamefont {Hillebrands}}, \bibinfo {author}
  {\bibfnamefont {B.}~\bibnamefont {Bartenlian}}, \bibinfo {author}
  {\bibfnamefont {C.}~\bibnamefont {Chappert}}, \bibinfo {author}
  {\bibfnamefont {F.}~\bibnamefont {Rousseaux}}, \ and\ \bibinfo {author}
  {\bibfnamefont {A.}~\bibnamefont {Slavin}},\ }\href {\doibase
  10.1103/PhysRevB.60.15194} {\bibfield  {journal} {\bibinfo  {journal}
  {Physical Review B}\ }\textbf {\bibinfo {volume} {60}},\ \bibinfo {pages}
  {15194} (\bibinfo {year} {1999})}\BibitemShut {NoStop}%
\bibitem [{\citenamefont {Buess}\ \emph {et~al.}(2004)\citenamefont {Buess},
  \citenamefont {H\"{o}llinger}, \citenamefont {Haug}, \citenamefont
  {Perzlmaier}, \citenamefont {Krey}, \citenamefont {Pescia}, \citenamefont
  {Scheinfein}, \citenamefont {Weiss},\ and\ \citenamefont {Back}}]{Buess2004}%
  \BibitemOpen
  \bibfield  {author} {\bibinfo {author} {\bibfnamefont {M.}~\bibnamefont
  {Buess}}, \bibinfo {author} {\bibfnamefont {R.}~\bibnamefont
  {H\"{o}llinger}}, \bibinfo {author} {\bibfnamefont {T.}~\bibnamefont {Haug}},
  \bibinfo {author} {\bibfnamefont {K.}~\bibnamefont {Perzlmaier}}, \bibinfo
  {author} {\bibfnamefont {U.}~\bibnamefont {Krey}}, \bibinfo {author}
  {\bibfnamefont {D.}~\bibnamefont {Pescia}}, \bibinfo {author} {\bibfnamefont
  {M.}~\bibnamefont {Scheinfein}}, \bibinfo {author} {\bibfnamefont
  {D.}~\bibnamefont {Weiss}}, \ and\ \bibinfo {author} {\bibfnamefont
  {C.}~\bibnamefont {Back}},\ }\href {\doibase 10.1103/PhysRevLett.93.077207}
  {\bibfield  {journal} {\bibinfo  {journal} {Physical Review Letters}\
  }\textbf {\bibinfo {volume} {93}},\ \bibinfo {pages} {077207} (\bibinfo
  {year} {2004})}\BibitemShut {NoStop}%
\bibitem [{\citenamefont {Bayer}\ \emph {et~al.}(2005)\citenamefont {Bayer},
  \citenamefont {Jorzick}, \citenamefont {Hillebrands}, \citenamefont
  {Demokritov}, \citenamefont {Kouba}, \citenamefont {Bozinoski}, \citenamefont
  {Slavin}, \citenamefont {Guslienko}, \citenamefont {Berkov}, \citenamefont
  {Gorn},\ and\ \citenamefont {Kostylev}}]{Bayer2005}%
  \BibitemOpen
  \bibfield  {author} {\bibinfo {author} {\bibfnamefont {C.}~\bibnamefont
  {Bayer}}, \bibinfo {author} {\bibfnamefont {J.}~\bibnamefont {Jorzick}},
  \bibinfo {author} {\bibfnamefont {B.}~\bibnamefont {Hillebrands}}, \bibinfo
  {author} {\bibfnamefont {S.}~\bibnamefont {Demokritov}}, \bibinfo {author}
  {\bibfnamefont {R.}~\bibnamefont {Kouba}}, \bibinfo {author} {\bibfnamefont
  {R.}~\bibnamefont {Bozinoski}}, \bibinfo {author} {\bibfnamefont
  {A.}~\bibnamefont {Slavin}}, \bibinfo {author} {\bibfnamefont
  {K.}~\bibnamefont {Guslienko}}, \bibinfo {author} {\bibfnamefont
  {D.}~\bibnamefont {Berkov}}, \bibinfo {author} {\bibfnamefont
  {N.}~\bibnamefont {Gorn}}, \ and\ \bibinfo {author} {\bibfnamefont
  {M.}~\bibnamefont {Kostylev}},\ }\href {\doibase 10.1103/PhysRevB.72.064427}
  {\bibfield  {journal} {\bibinfo  {journal} {Physical Review B}\ }\textbf
  {\bibinfo {volume} {72}},\ \bibinfo {pages} {064427} (\bibinfo {year}
  {2005})}\BibitemShut {NoStop}%
\bibitem [{\citenamefont {Perzlmaier}\ \emph {et~al.}(2005)\citenamefont
  {Perzlmaier}, \citenamefont {Buess}, \citenamefont {Back}, \citenamefont
  {Demidov}, \citenamefont {Hillebrands},\ and\ \citenamefont
  {Demokritov}}]{Perzlmaier2005}%
  \BibitemOpen
  \bibfield  {author} {\bibinfo {author} {\bibfnamefont {K.}~\bibnamefont
  {Perzlmaier}}, \bibinfo {author} {\bibfnamefont {M.}~\bibnamefont {Buess}},
  \bibinfo {author} {\bibfnamefont {C.}~\bibnamefont {Back}}, \bibinfo {author}
  {\bibfnamefont {V.}~\bibnamefont {Demidov}}, \bibinfo {author} {\bibfnamefont
  {B.}~\bibnamefont {Hillebrands}}, \ and\ \bibinfo {author} {\bibfnamefont
  {S.}~\bibnamefont {Demokritov}},\ }\href {\doibase
  10.1103/PhysRevLett.94.057202} {\bibfield  {journal} {\bibinfo  {journal}
  {Physical Review Letters}\ }\textbf {\bibinfo {volume} {94}},\ \bibinfo
  {pages} {057202} (\bibinfo {year} {2005})}\BibitemShut {NoStop}%
\bibitem [{\citenamefont {Neudecker}\ \emph {et~al.}(2006)\citenamefont
  {Neudecker}, \citenamefont {Perzlmaier}, \citenamefont {Hoffmann},
  \citenamefont {Woltersdorf}, \citenamefont {Buess}, \citenamefont {Weiss},\
  and\ \citenamefont {Back}}]{Neudecker2006a}%
  \BibitemOpen
  \bibfield  {author} {\bibinfo {author} {\bibfnamefont {I.}~\bibnamefont
  {Neudecker}}, \bibinfo {author} {\bibfnamefont {K.}~\bibnamefont
  {Perzlmaier}}, \bibinfo {author} {\bibfnamefont {F.}~\bibnamefont
  {Hoffmann}}, \bibinfo {author} {\bibfnamefont {G.}~\bibnamefont
  {Woltersdorf}}, \bibinfo {author} {\bibfnamefont {M.}~\bibnamefont {Buess}},
  \bibinfo {author} {\bibfnamefont {D.}~\bibnamefont {Weiss}}, \ and\ \bibinfo
  {author} {\bibfnamefont {C.}~\bibnamefont {Back}},\ }\href {\doibase
  10.1103/PhysRevB.73.134426} {\bibfield  {journal} {\bibinfo  {journal}
  {Physical Review B}\ }\textbf {\bibinfo {volume} {73}},\ \bibinfo {pages}
  {134426} (\bibinfo {year} {2006})}\BibitemShut {NoStop}%
\bibitem [{\citenamefont {Podbielski}\ \emph {et~al.}(2006)\citenamefont
  {Podbielski}, \citenamefont {Giesen},\ and\ \citenamefont
  {Grundler}}]{Podbielski2006}%
  \BibitemOpen
  \bibfield  {author} {\bibinfo {author} {\bibfnamefont {J.}~\bibnamefont
  {Podbielski}}, \bibinfo {author} {\bibfnamefont {F.}~\bibnamefont {Giesen}},
  \ and\ \bibinfo {author} {\bibfnamefont {D.}~\bibnamefont {Grundler}},\
  }\href {\doibase 10.1103/PhysRevLett.96.167207} {\bibfield  {journal}
  {\bibinfo  {journal} {Physical Review Letters}\ }\textbf {\bibinfo {volume}
  {96}},\ \bibinfo {pages} {167207} (\bibinfo {year} {2006})}\BibitemShut
  {NoStop}%
\bibitem [{\citenamefont {Mendach}\ \emph {et~al.}(2008)\citenamefont
  {Mendach}, \citenamefont {Podbielski}, \citenamefont {Topp}, \citenamefont
  {Hansen},\ and\ \citenamefont {Heitmann}}]{Mendach2008}%
  \BibitemOpen
  \bibfield  {author} {\bibinfo {author} {\bibfnamefont {S.}~\bibnamefont
  {Mendach}}, \bibinfo {author} {\bibfnamefont {J.}~\bibnamefont {Podbielski}},
  \bibinfo {author} {\bibfnamefont {J.}~\bibnamefont {Topp}}, \bibinfo {author}
  {\bibfnamefont {W.}~\bibnamefont {Hansen}}, \ and\ \bibinfo {author}
  {\bibfnamefont {D.}~\bibnamefont {Heitmann}},\ }\href {\doibase
  10.1063/1.3058764} {\bibfield  {journal} {\bibinfo  {journal} {Applied
  Physics Letters}\ }\textbf {\bibinfo {volume} {93}},\ \bibinfo {pages}
  {262501} (\bibinfo {year} {2008})}\BibitemShut {NoStop}%
\bibitem [{\citenamefont {Balhorn}\ \emph {et~al.}(2010)\citenamefont
  {Balhorn}, \citenamefont {Mansfeld}, \citenamefont {Krohn}, \citenamefont
  {Topp}, \citenamefont {Hansen}, \citenamefont {Heitmann},\ and\ \citenamefont
  {Mendach}}]{Balhorn2010}%
  \BibitemOpen
  \bibfield  {author} {\bibinfo {author} {\bibfnamefont {F.}~\bibnamefont
  {Balhorn}}, \bibinfo {author} {\bibfnamefont {S.}~\bibnamefont {Mansfeld}},
  \bibinfo {author} {\bibfnamefont {A.}~\bibnamefont {Krohn}}, \bibinfo
  {author} {\bibfnamefont {J.}~\bibnamefont {Topp}}, \bibinfo {author}
  {\bibfnamefont {W.}~\bibnamefont {Hansen}}, \bibinfo {author} {\bibfnamefont
  {D.}~\bibnamefont {Heitmann}}, \ and\ \bibinfo {author} {\bibfnamefont
  {S.}~\bibnamefont {Mendach}},\ }\href {\doibase
  10.1103/PhysRevLett.104.037205} {\bibfield  {journal} {\bibinfo  {journal}
  {Physical Review Letters}\ }\textbf {\bibinfo {volume} {104}},\ \bibinfo
  {pages} {037205} (\bibinfo {year} {2010})}\BibitemShut {NoStop}%
\bibitem [{\citenamefont {Bauer}\ \emph {et~al.}(1997)\citenamefont {Bauer},
  \citenamefont {Mathieu}, \citenamefont {Demokritov}, \citenamefont
  {Hillebrands}, \citenamefont {Kolodin}, \citenamefont {Sure}, \citenamefont
  {D\"otsch}, \citenamefont {Grimalsky}, \citenamefont {Rapoport},\ and\
  \citenamefont {Slavin}}]{Hillebrands1997}%
  \BibitemOpen
  \bibfield  {author} {\bibinfo {author} {\bibfnamefont {M.}~\bibnamefont
  {Bauer}}, \bibinfo {author} {\bibfnamefont {C.}~\bibnamefont {Mathieu}},
  \bibinfo {author} {\bibfnamefont {S.~O.}\ \bibnamefont {Demokritov}},
  \bibinfo {author} {\bibfnamefont {B.}~\bibnamefont {Hillebrands}}, \bibinfo
  {author} {\bibfnamefont {P.~A.}\ \bibnamefont {Kolodin}}, \bibinfo {author}
  {\bibfnamefont {S.}~\bibnamefont {Sure}}, \bibinfo {author} {\bibfnamefont
  {H.}~\bibnamefont {D\"otsch}}, \bibinfo {author} {\bibfnamefont
  {V.}~\bibnamefont {Grimalsky}}, \bibinfo {author} {\bibfnamefont
  {Y.}~\bibnamefont {Rapoport}}, \ and\ \bibinfo {author} {\bibfnamefont
  {A.~N.}\ \bibnamefont {Slavin}},\ }\href {\doibase 10.1103/PhysRevB.56.R8483}
  {\bibfield  {journal} {\bibinfo  {journal} {Phys. Rev. B}\ }\textbf {\bibinfo
  {volume} {56}},\ \bibinfo {pages} {R8483} (\bibinfo {year}
  {1997})}\BibitemShut {NoStop}%
\bibitem [{\citenamefont {Bauer}\ \emph {et~al.}(1998)\citenamefont {Bauer},
  \citenamefont {B\"{u}ttner}, \citenamefont {Demokritov}, \citenamefont
  {Hillebrands}, \citenamefont {Grimalsky}, \citenamefont {Rapoport},\ and\
  \citenamefont {Slavin}}]{Bauer1998}%
  \BibitemOpen
  \bibfield  {author} {\bibinfo {author} {\bibfnamefont {M.}~\bibnamefont
  {Bauer}}, \bibinfo {author} {\bibfnamefont {O.}~\bibnamefont {B\"{u}ttner}},
  \bibinfo {author} {\bibfnamefont {S.}~\bibnamefont {Demokritov}}, \bibinfo
  {author} {\bibfnamefont {B.}~\bibnamefont {Hillebrands}}, \bibinfo {author}
  {\bibfnamefont {V.}~\bibnamefont {Grimalsky}}, \bibinfo {author}
  {\bibfnamefont {Y.}~\bibnamefont {Rapoport}}, \ and\ \bibinfo {author}
  {\bibfnamefont {A.}~\bibnamefont {Slavin}},\ }\href {\doibase
  10.1103/PhysRevLett.81.3769} {\bibfield  {journal} {\bibinfo  {journal}
  {Physical Review Letters}\ }\textbf {\bibinfo {volume} {81}},\ \bibinfo
  {pages} {3769} (\bibinfo {year} {1998})}\BibitemShut {NoStop}%
\bibitem [{\citenamefont {Demidov}\ \emph {et~al.}(2007)\citenamefont
  {Demidov}, \citenamefont {Demokritov}, \citenamefont {Rott}, \citenamefont
  {Krzysteczko},\ and\ \citenamefont {Reiss}}]{Demidov2007}%
  \BibitemOpen
  \bibfield  {author} {\bibinfo {author} {\bibfnamefont {V.~E.}\ \bibnamefont
  {Demidov}}, \bibinfo {author} {\bibfnamefont {S.~O.}\ \bibnamefont
  {Demokritov}}, \bibinfo {author} {\bibfnamefont {K.}~\bibnamefont {Rott}},
  \bibinfo {author} {\bibfnamefont {P.}~\bibnamefont {Krzysteczko}}, \ and\
  \bibinfo {author} {\bibfnamefont {G.}~\bibnamefont {Reiss}},\ }\href
  {\doibase 10.1063/1.2825421} {\bibfield  {journal} {\bibinfo  {journal}
  {Applied Physics Letters}\ }\textbf {\bibinfo {volume} {91}},\ \bibinfo
  {pages} {252504} (\bibinfo {year} {2007})}\BibitemShut {NoStop}%
\bibitem [{\citenamefont {Topp}\ \emph {et~al.}(2008)\citenamefont {Topp},
  \citenamefont {Podbielski}, \citenamefont {Heitmann},\ and\ \citenamefont
  {Grundler}}]{Topp2008}%
  \BibitemOpen
  \bibfield  {author} {\bibinfo {author} {\bibfnamefont {J.}~\bibnamefont
  {Topp}}, \bibinfo {author} {\bibfnamefont {J.}~\bibnamefont {Podbielski}},
  \bibinfo {author} {\bibfnamefont {D.}~\bibnamefont {Heitmann}}, \ and\
  \bibinfo {author} {\bibfnamefont {D.}~\bibnamefont {Grundler}},\ }\href
  {\doibase 10.1103/PhysRevB.78.024431} {\bibfield  {journal} {\bibinfo
  {journal} {Physical Review B}\ }\textbf {\bibinfo {volume} {78}},\ \bibinfo
  {pages} {024431} (\bibinfo {year} {2008})}\BibitemShut {NoStop}%
\bibitem [{\citenamefont {Demidov}\ \emph {et~al.}(2008)\citenamefont
  {Demidov}, \citenamefont {Demokritov}, \citenamefont {Rott}, \citenamefont
  {Krzysteczko},\ and\ \citenamefont {Reiss}}]{Demidov2008b}%
  \BibitemOpen
  \bibfield  {author} {\bibinfo {author} {\bibfnamefont {V.}~\bibnamefont
  {Demidov}}, \bibinfo {author} {\bibfnamefont {S.}~\bibnamefont {Demokritov}},
  \bibinfo {author} {\bibfnamefont {K.}~\bibnamefont {Rott}}, \bibinfo {author}
  {\bibfnamefont {P.}~\bibnamefont {Krzysteczko}}, \ and\ \bibinfo {author}
  {\bibfnamefont {G.}~\bibnamefont {Reiss}},\ }\href {\doibase
  10.1103/PhysRevB.77.064406} {\bibfield  {journal} {\bibinfo  {journal}
  {Physical Review B}\ }\textbf {\bibinfo {volume} {77}},\ \bibinfo {pages}
  {064406} (\bibinfo {year} {2008})}\BibitemShut {NoStop}%
\bibitem [{\citenamefont {Demidov}\ \emph
  {et~al.}(2009{\natexlab{a}})\citenamefont {Demidov}, \citenamefont {Jersch},
  \citenamefont {Demokritov}, \citenamefont {Rott}, \citenamefont
  {Krzysteczko},\ and\ \citenamefont {Reiss}}]{Demidov2009c}%
  \BibitemOpen
  \bibfield  {author} {\bibinfo {author} {\bibfnamefont {V.}~\bibnamefont
  {Demidov}}, \bibinfo {author} {\bibfnamefont {J.}~\bibnamefont {Jersch}},
  \bibinfo {author} {\bibfnamefont {S.}~\bibnamefont {Demokritov}}, \bibinfo
  {author} {\bibfnamefont {K.}~\bibnamefont {Rott}}, \bibinfo {author}
  {\bibfnamefont {P.}~\bibnamefont {Krzysteczko}}, \ and\ \bibinfo {author}
  {\bibfnamefont {G.}~\bibnamefont {Reiss}},\ }\href {\doibase
  10.1103/PhysRevB.79.054417} {\bibfield  {journal} {\bibinfo  {journal}
  {Physical Review B}\ }\textbf {\bibinfo {volume} {79}},\ \bibinfo {pages}
  {054417} (\bibinfo {year} {2009}{\natexlab{a}})}\BibitemShut {NoStop}%
\bibitem [{\citenamefont {B\"uttner}\ \emph {et~al.}(2000)\citenamefont
  {B\"uttner}, \citenamefont {Bauer}, \citenamefont {Demokritov}, \citenamefont
  {Hillebrands}, \citenamefont {Kivshar}, \citenamefont {Grimalsky},
  \citenamefont {Rapoport},\ and\ \citenamefont {Slavin}}]{Buettner2000}%
  \BibitemOpen
  \bibfield  {author} {\bibinfo {author} {\bibfnamefont {O.}~\bibnamefont
  {B\"uttner}}, \bibinfo {author} {\bibfnamefont {M.}~\bibnamefont {Bauer}},
  \bibinfo {author} {\bibfnamefont {S.~O.}\ \bibnamefont {Demokritov}},
  \bibinfo {author} {\bibfnamefont {B.}~\bibnamefont {Hillebrands}}, \bibinfo
  {author} {\bibfnamefont {Y.~S.}\ \bibnamefont {Kivshar}}, \bibinfo {author}
  {\bibfnamefont {V.}~\bibnamefont {Grimalsky}}, \bibinfo {author}
  {\bibfnamefont {Y.}~\bibnamefont {Rapoport}}, \ and\ \bibinfo {author}
  {\bibfnamefont {A.~N.}\ \bibnamefont {Slavin}},\ }\href {\doibase
  10.1103/PhysRevB.61.11576} {\bibfield  {journal} {\bibinfo  {journal} {Phys.
  Rev. B}\ }\textbf {\bibinfo {volume} {61}},\ \bibinfo {pages} {11576}
  (\bibinfo {year} {2000})}\BibitemShut {NoStop}%
\bibitem [{\citenamefont {Demidov}\ \emph {et~al.}(2005)\citenamefont
  {Demidov}, \citenamefont {Hillebrands}, \citenamefont {Demokritov},
  \citenamefont {Laufenberg},\ and\ \citenamefont {Freitas}}]{Demidov2005}%
  \BibitemOpen
  \bibfield  {author} {\bibinfo {author} {\bibfnamefont {V.~E.}\ \bibnamefont
  {Demidov}}, \bibinfo {author} {\bibfnamefont {B.}~\bibnamefont
  {Hillebrands}}, \bibinfo {author} {\bibfnamefont {S.~O.}\ \bibnamefont
  {Demokritov}}, \bibinfo {author} {\bibfnamefont {M.}~\bibnamefont
  {Laufenberg}}, \ and\ \bibinfo {author} {\bibfnamefont {P.~P.}\ \bibnamefont
  {Freitas}},\ }\href {\doibase DOI:10.1063/1.1855014} {\bibfield  {journal}
  {\bibinfo  {journal} {Journal of Applied Physics}\ }\textbf {\bibinfo
  {volume} {97}},\ \bibinfo {pages} {10A717} (\bibinfo {year}
  {2005})}\BibitemShut {NoStop}%
\bibitem [{\citenamefont {Perzlmaier}\ \emph {et~al.}(2008)\citenamefont
  {Perzlmaier}, \citenamefont {Woltersdorf},\ and\ \citenamefont
  {Back}}]{Perzlmaier2008}%
  \BibitemOpen
  \bibfield  {author} {\bibinfo {author} {\bibfnamefont {K.}~\bibnamefont
  {Perzlmaier}}, \bibinfo {author} {\bibfnamefont {G.}~\bibnamefont
  {Woltersdorf}}, \ and\ \bibinfo {author} {\bibfnamefont {C.}~\bibnamefont
  {Back}},\ }\href {\doibase 10.1103/PhysRevB.77.054425} {\bibfield  {journal}
  {\bibinfo  {journal} {Physical Review B}\ }\textbf {\bibinfo {volume} {77}},\
  \bibinfo {pages} {054425} (\bibinfo {year} {2008})}\BibitemShut {NoStop}%
\bibitem [{\citenamefont {Veerakumar}\ and\ \citenamefont
  {Camley}(2006)}]{Veerakumar2006}%
  \BibitemOpen
  \bibfield  {author} {\bibinfo {author} {\bibfnamefont {V.}~\bibnamefont
  {Veerakumar}}\ and\ \bibinfo {author} {\bibfnamefont {R.~E.}\ \bibnamefont
  {Camley}},\ }\href {\doibase 10.1103/PhysRevB.74.214401} {\bibfield
  {journal} {\bibinfo  {journal} {Phys. Rev. B}\ }\textbf {\bibinfo {volume}
  {74}},\ \bibinfo {pages} {214401} (\bibinfo {year} {2006})}\BibitemShut
  {NoStop}%
\bibitem [{\citenamefont {Demidov}\ \emph
  {et~al.}(2009{\natexlab{b}})\citenamefont {Demidov}, \citenamefont
  {Demokritov}, \citenamefont {Birt}, \citenamefont {O'Gorman}, \citenamefont
  {Tsoi},\ and\ \citenamefont {Li}}]{Demidov2009d}%
  \BibitemOpen
  \bibfield  {author} {\bibinfo {author} {\bibfnamefont {V.~E.}\ \bibnamefont
  {Demidov}}, \bibinfo {author} {\bibfnamefont {S.~O.}\ \bibnamefont
  {Demokritov}}, \bibinfo {author} {\bibfnamefont {D.}~\bibnamefont {Birt}},
  \bibinfo {author} {\bibfnamefont {B.}~\bibnamefont {O'Gorman}}, \bibinfo
  {author} {\bibfnamefont {M.}~\bibnamefont {Tsoi}}, \ and\ \bibinfo {author}
  {\bibfnamefont {X.}~\bibnamefont {Li}},\ }\href {\doibase
  10.1103/PhysRevB.80.014429} {\bibfield  {journal} {\bibinfo  {journal} {Phys.
  Rev. B}\ }\textbf {\bibinfo {volume} {80}},\ \bibinfo {pages} {014429}
  (\bibinfo {year} {2009}{\natexlab{b}})}\BibitemShut {NoStop}%
\bibitem [{\citenamefont {Schneider}\ \emph {et~al.}(2010)\citenamefont
  {Schneider}, \citenamefont {Serga}, \citenamefont {Chumak}, \citenamefont
  {Sandweg}, \citenamefont {Trudel}, \citenamefont {Wolff}, \citenamefont
  {Kostylev}, \citenamefont {Tiberkevich}, \citenamefont {Slavin},\ and\
  \citenamefont {Hillebrands}}]{Schneider2010}%
  \BibitemOpen
  \bibfield  {author} {\bibinfo {author} {\bibfnamefont {T.}~\bibnamefont
  {Schneider}}, \bibinfo {author} {\bibfnamefont {A.~A.}\ \bibnamefont
  {Serga}}, \bibinfo {author} {\bibfnamefont {A.~V.}\ \bibnamefont {Chumak}},
  \bibinfo {author} {\bibfnamefont {C.~W.}\ \bibnamefont {Sandweg}}, \bibinfo
  {author} {\bibfnamefont {S.}~\bibnamefont {Trudel}}, \bibinfo {author}
  {\bibfnamefont {S.}~\bibnamefont {Wolff}}, \bibinfo {author} {\bibfnamefont
  {M.~P.}\ \bibnamefont {Kostylev}}, \bibinfo {author} {\bibfnamefont {V.~S.}\
  \bibnamefont {Tiberkevich}}, \bibinfo {author} {\bibfnamefont {A.~N.}\
  \bibnamefont {Slavin}}, \ and\ \bibinfo {author} {\bibfnamefont
  {B.}~\bibnamefont {Hillebrands}},\ }\href {\doibase
  10.1103/PhysRevLett.104.197203} {\bibfield  {journal} {\bibinfo  {journal}
  {Phys. Rev. Lett.}\ }\textbf {\bibinfo {volume} {104}},\ \bibinfo {pages}
  {197203} (\bibinfo {year} {2010})}\BibitemShut {NoStop}%
\bibitem [{\citenamefont {Kostylev}\ and\ \citenamefont
  {Hillebrands}(2011)}]{Kostylev2011}%
  \BibitemOpen
  \bibfield  {author} {\bibinfo {author} {\bibfnamefont {A.~A.}\ \bibnamefont
  {Kostylev}, \bibfnamefont {M.~P.~Serga}}\ and\ \bibinfo {author}
  {\bibfnamefont {B.}~\bibnamefont {Hillebrands}},\ }\href@noop {} {\bibfield
  {journal} {\bibinfo  {journal} {Physics Review Letters}\ }\textbf {\bibinfo
  {volume} {106}},\ \bibinfo {pages} {134101} (\bibinfo {year}
  {2011})}\BibitemShut {NoStop}%
\bibitem [{\citenamefont {Freeman}\ and\ \citenamefont
  {Smyth}(1996)}]{Freeman1996}%
  \BibitemOpen
  \bibfield  {author} {\bibinfo {author} {\bibfnamefont {M.~R.}\ \bibnamefont
  {Freeman}}\ and\ \bibinfo {author} {\bibfnamefont {J.~F.}\ \bibnamefont
  {Smyth}},\ }\href {\doibase DOI:10.1063/1.361896} {\bibfield  {journal}
  {\bibinfo  {journal} {Journal of Applied Physics}\ }\textbf {\bibinfo
  {volume} {79}},\ \bibinfo {pages} {5898} (\bibinfo {year}
  {1996})}\BibitemShut {NoStop}%
\bibitem [{\citenamefont {Tamaru}\ \emph {et~al.}(2004)\citenamefont {Tamaru},
  \citenamefont {Bain}, \citenamefont {{van De Veerdonk}}, \citenamefont
  {Crawford}, \citenamefont {Covington},\ and\ \citenamefont
  {Kryder}}]{Tamaru2004}%
  \BibitemOpen
  \bibfield  {author} {\bibinfo {author} {\bibfnamefont {S.}~\bibnamefont
  {Tamaru}}, \bibinfo {author} {\bibfnamefont {J.}~\bibnamefont {Bain}},
  \bibinfo {author} {\bibfnamefont {R.}~\bibnamefont {{van De Veerdonk}}},
  \bibinfo {author} {\bibfnamefont {T.}~\bibnamefont {Crawford}}, \bibinfo
  {author} {\bibfnamefont {M.}~\bibnamefont {Covington}}, \ and\ \bibinfo
  {author} {\bibfnamefont {M.}~\bibnamefont {Kryder}},\ }\href {\doibase
  10.1103/PhysRevB.70.104416} {\bibfield  {journal} {\bibinfo  {journal}
  {Physical Review B}\ }\textbf {\bibinfo {volume} {70}},\ \bibinfo {pages}
  {104416} (\bibinfo {year} {2004})}\BibitemShut {NoStop}%
\bibitem [{\citenamefont {Abbe}(1873)}]{Abbe1873}%
  \BibitemOpen
  \bibfield  {author} {\bibinfo {author} {\bibfnamefont {E.~K.}\ \bibnamefont
  {Abbe}},\ }\href@noop {} {\bibfield  {journal} {\bibinfo  {journal} {Archiv
  fuer Mikroskopische Anatomie}\ }\textbf {\bibinfo {volume} {9}},\ \bibinfo
  {pages} {413} (\bibinfo {year} {1873})}\BibitemShut {NoStop}%
\bibitem [{\citenamefont {Liu}\ and\ \citenamefont {Zhang}(2007)}]{Liu2007b}%
  \BibitemOpen
  \bibfield  {author} {\bibinfo {author} {\bibfnamefont {H.~X. Y. S.~C.}\
  \bibnamefont {Liu}, \bibfnamefont {Z.~Lee}}\ and\ \bibinfo {author}
  {\bibfnamefont {X.}~\bibnamefont {Zhang}},\ }\href@noop {} {\bibfield
  {journal} {\bibinfo  {journal} {Science}\ }\textbf {\bibinfo {volume}
  {315}},\ \bibinfo {pages} {1686} (\bibinfo {year} {2007})}\BibitemShut
  {NoStop}%
\bibitem [{\citenamefont {Schwaiger}\ \emph {et~al.}(2009)\citenamefont
  {Schwaiger}, \citenamefont {Br\"{o}ll}, \citenamefont {Krohn}, \citenamefont
  {Stemmann}, \citenamefont {Heyn}, \citenamefont {Stark}, \citenamefont
  {Stickler}, \citenamefont {Heitmann},\ and\ \citenamefont
  {Mendach}}]{Schwaiger2009}%
  \BibitemOpen
  \bibfield  {author} {\bibinfo {author} {\bibfnamefont {S.}~\bibnamefont
  {Schwaiger}}, \bibinfo {author} {\bibfnamefont {M.}~\bibnamefont
  {Br\"{o}ll}}, \bibinfo {author} {\bibfnamefont {A.}~\bibnamefont {Krohn}},
  \bibinfo {author} {\bibfnamefont {A.}~\bibnamefont {Stemmann}}, \bibinfo
  {author} {\bibfnamefont {C.}~\bibnamefont {Heyn}}, \bibinfo {author}
  {\bibfnamefont {Y.}~\bibnamefont {Stark}}, \bibinfo {author} {\bibfnamefont
  {D.}~\bibnamefont {Stickler}}, \bibinfo {author} {\bibfnamefont
  {D.}~\bibnamefont {Heitmann}}, \ and\ \bibinfo {author} {\bibfnamefont
  {S.}~\bibnamefont {Mendach}},\ }\href {\doibase
  10.1103/PhysRevLett.102.163903} {\bibfield  {journal} {\bibinfo  {journal}
  {Physical Review Letters}\ }\textbf {\bibinfo {volume} {102}},\ \bibinfo
  {pages} {163903} (\bibinfo {year} {2009})}\BibitemShut {NoStop}%
\bibitem [{\citenamefont {Guslienko}\ \emph {et~al.}(2003)\citenamefont
  {Guslienko}, \citenamefont {Chantrell},\ and\ \citenamefont
  {Slavin}}]{Guslienko2003}%
  \BibitemOpen
  \bibfield  {author} {\bibinfo {author} {\bibfnamefont {K.}~\bibnamefont
  {Guslienko}}, \bibinfo {author} {\bibfnamefont {R.}~\bibnamefont
  {Chantrell}}, \ and\ \bibinfo {author} {\bibfnamefont {A.}~\bibnamefont
  {Slavin}},\ }\href {\doibase 10.1103/PhysRevB.68.024422} {\bibfield
  {journal} {\bibinfo  {journal} {Physical Review B}\ }\textbf {\bibinfo
  {volume} {68}},\ \bibinfo {pages} {024422} (\bibinfo {year}
  {2003})}\BibitemShut {NoStop}%
\end{thebibliography}
\end{document}